\documentclass{llncs}
\usepackage{tabularx}
\usepackage{xcolor}
\usepackage{graphicx,float,subfig,epstopdf}
\usepackage{url}
\begin{document}\newcommand{\enzo}[1]{{\textcolor{red}{[#1]}}}

\title{Anatomical Priors for Image Segmentation via Post-Processing with Denoising Autoencoders}

\author{Agostina J. Larrazabal, Cesar Martinez, Enzo Ferrante}
\institute{Research institute for signals, systems and computational intelligence, sinc(i), FICH-UNL / CONICET, Santa Fe, Argentina}

\titlerunning{Anatomical Priors for Image Segmentation via Post-Processing with Denoising
  Recompile
1
 Autoencoders}
% abbreviated title (for running head)
\maketitle

\begin{abstract}
Deep convolutional neural networks (CNN) proved to be highly accurate to perform anatomical segmentation of medical images. However, some of the most popular CNN architectures for image segmentation still rely on post-processing strategies (e.g. Conditional Random Fields) to incorporate connectivity constraints into the resulting masks. These post-processing steps are based on the assumption that objects are usually continuous and therefore nearby pixels should be assigned the same object label. Even if it is a valid assumption in general, these methods do not offer a straightforward way to incorporate more complex priors like convexity or arbitrary shape restrictions.\\ %Moreover, they usually increase the overall time complexity of the complete segmentation pipeline.
\indent In this work we propose Post-DAE, a post-processing method based on denoising autoencoders (DAE) trained using only segmentation masks. We learn a low-dimensional space of anatomically plausible segmentations, and use it as a post-processing step to impose shape constraints on the resulting masks obtained with arbitrary segmentation methods. Our approach is independent of image modality and intensity information since it employs only segmentation masks for training. This enables the use of anatomical segmentations that do not need to be paired with intensity images, making the approach very flexible. Our experimental results on anatomical segmentation of X-ray images show that Post-DAE can improve the quality of noisy and incorrect segmentation masks obtained with a variety of standard methods, by bringing them back to a feasible space, with almost no extra computational time.

\end{abstract}
\section{Introduction}
Segmentation of anatomical structures is a fundamental task for biomedical image analysis. It constitutes the first step in several medical procedures such as shape analysis for population studies, computed assisted diagnosis and automatic radiotherapy planning, among many others. The accuracy and anatomical plausibility of these segmentations is therefore of paramount importance, since it will necessarily influence the overall quality of such procedures.

\begin{figure}[t!]
   \includegraphics[width=\textwidth]{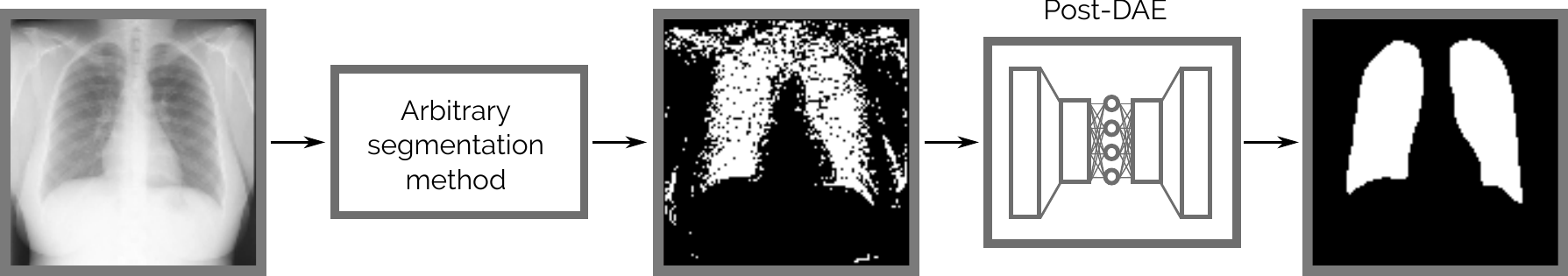}
    \caption{Post-DAE works as a post-processing step and improves the anatomical plausability of segmentation masks obtained with arbitrary methods.}
\label{fig:workflow}   
\end{figure}

During the last years, convolutional neural networks (CNNs) proved to be highly accurate to perform segmentation in biomedical images \cite{RonnebergerUnet15,KamnitsasDeepmedic16,Shakeri2016}. 
%CNNs constitute a particular type of neural networks specially suited for regularly structured data, like 2D or 3D images, where multiple representations of the input are learned at every successive layer.
One of the tricks that enables the use of CNNs in large images (by reducing the number of learned parameters) is known as parameter sharing scheme. The assumption behind this idea is that, at every layer, shared parameters are used to learn new representations of the input data along the whole image. These parameters (also referred as weights or kernels) are successively convoluted with the input data resulting in more abstract representations. This trick is especially useful for tasks like image classification, where invariance to translation is a desired property since objects may appear in any location. However, in case of anatomical structures in medical images where their location tend to be highly regular, this property leads to incorrect predictions in areas with similar intensities when enough contextual information is not considered. Shape and topology tend also to be preserved in anatomical images of the same type. However, as discussed in \cite{BenTaieb2016}, the pixel-level predictions of most CNN architectures are not designed to account for higher-order topological properties. 

Before the advent of CNNs, other classical learning based segmentation methods were popular for this task (e.g.  Random Forest (RF) \cite{breiman2001random}), some of which are still being used specially when the amount of annotated data is not enough to train deep CNNs. The pixel-level predictions of these approaches are also influenced by image patches of fixed size. In these cases, handcrafted features are extracted from image patches and used to train a classifier, which predicts the class corresponding to the central pixel in that patch. These methods suffer from the same limitations related to the lack of shape and topological information discussed before.

In this work, we introduce Post-DAE (post-processing with denoising autoencoders), a post-processing method which produces anatomically plausible segmentations by improving pixel-level predictions coming from arbitrary classifiers (e.g. CNNs or RF), incorporating shape and topological priors. We employ Denoising Autoencoders (DAE) to learn compact and non-linear representations of anatomical structures, using only segmentation masks. This model is then applied as a post-processing method for image segmentation, bringing arbitrary and potentially erroneous segmentation masks into an anatomically plausible space (see Figure \ref{fig:workflow}). \\

\noindent \textbf{Contributions.} Our contributions are 3-fold: (i) we show, for the first time, that DAE can be used as an independent post-processing step to correct problematic and non-anatomically plausible masks produced by arbitrary segmentation methods; (ii) we design a method that can be trained using segmentation-only datasets or anatomical masks coming from arbitrary image modalities, since the DAE is trained using only segmentation masks, and no intensity information is required during learning; (iii) we validate Post-DAE in the context of lung segmentation in X-ray images, bench-marking with other classical post-processing method and showing its robustness by improving segmentation masks coming from both, CNN and RF-based classifiers.\\%Our method produces substantial improvement in non-realistic segmentations, while refining the results when segmentations are already of acceptable quality.\\
%, enabling the flexibility of having an external description of the anatomy that need not be available in the current data

\noindent \textbf{Related works.} 
%Multiple alternatives have been proposed  (see \cite{nosrati2016incorporating} for a complete review).
%to alleviate the aforementioned issues in the context of medical images. 
%A simple but effective approach is to increase the receptive field of the network, i.e. the area of the input image that influences a single prediction. In this regard, \cite{KamnitsasDeepmedic16} proposed to increase the receptive field of the CNN by means of a dual path focusing on a wider low resolution area of the input image. This increases the contextual information provided to the network, but also augments the complexity of the segmentation model itself. To avoid extra computational cost, 
One popular strategy to incorporate prior knowledge about shape and topology into medical image segmentation is to modify the loss used to train the model. The work of \cite{BenTaieb2016} incorporates high-order regularization through a topology aware loss function. The main disadvantage is that such loss function is constructed ad-hoc for every dataset, requiring the user to manually specify the topological relations between the semantic classes through a topological validity table. More similar to our work are those by \cite{Oktay2017,Ravishankar2017}, where an autoencoder (AE) is used to learn lower dimensional representations of image anatomy. The AE is used to define a loss term that imposes anatomical constraints during training. The main disadvantage of these approaches is that they can only be used during training of CNN architectures. Other methods like RF-based segmentation can not be improved through this technique. On the contrary, our method post-processes arbitrary segmentation masks. Therefore, it can be used to improve results obtained with any segmentation method, even those methods which do not rely on an explicit training phase (e.g. level-sets methods).

Post-processing methods have also been considered in the literature. In \cite{Shakeri2016}, the output CNN scores are considered as unary potentials of a Markov random field (MRF) energy minimization problem, where spatial homogeneity is propagated through pairwise relations. Similarly, \cite{KamnitsasDeepmedic16} uses  a fully connected conditional random field (CRF) as post-processing step. However, as stated by \cite{KamnitsasDeepmedic16}, finding a global set of parameters for the graphical models which can consistently improve the segmentation of all classes remains a challenging problem. Moreover, these methods do not incorporate shape priors. Instead, they are based on the assumption that objects are usually continuous and therefore nearby pixels (or pixels with similar appearence) should be assigned the same object label. Conversely, our post-processing method makes use of a DAE to impose shape priors, transforming any segmentation mask into an anatomically plausible one.
\section{Anatomical Priors for Image Segmentation via Post-Processing with DAE}
\begin{figure}[t!]
   \includegraphics[width=\textwidth]{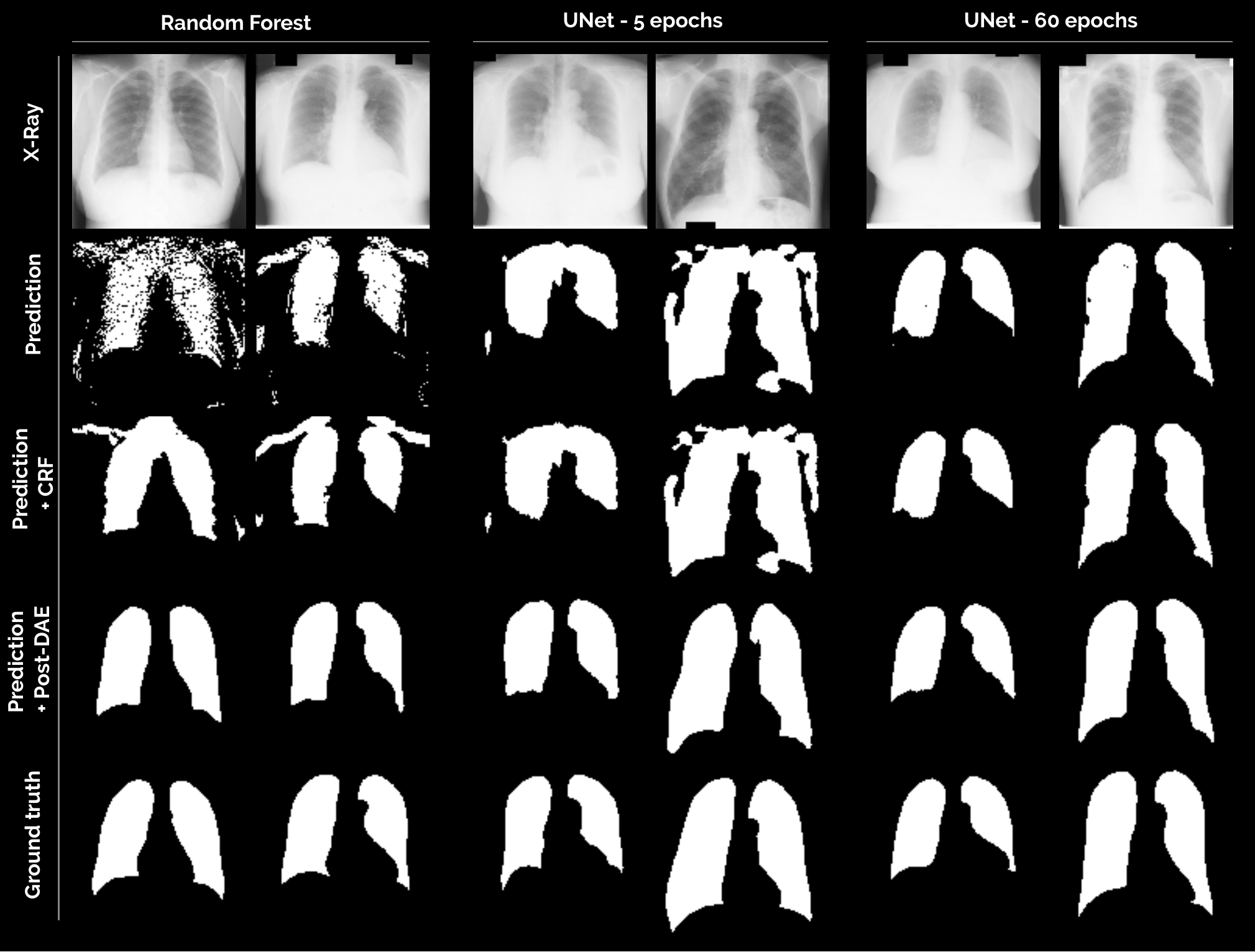}
    \caption{Predictions obtained with three different methods: Random Forest, UNet trained for 5 epochs and until convergence. Rows from top to bottom: (i) X-Ray image; (ii) original mask obtained with the corresponding method; (iii) mask post-processed with a fully connected CRF; (iv) mask post-processed with the proposed Post-DAE method; and (v) ground-truth. }
\label{fig:qualitativeResults}   
\end{figure}
\noindent \textbf{Problem statement.} Given a dataset of unpaired anatomical segmentation masks $\mathcal{D_A} = \{S^A_i\}_{0 \leq i \leq |\mathcal{D_A}|}$ (unpaired in the sense that no corresponding intensity image associated to the segmentation mask is required) we aim at learning a model that can bring segmentations $\mathcal{D_P} = \{S^P_i\}_{0 \leq i \leq |\mathcal{D_P}|}$ predicted by arbitrary classifiers $P$ into an anatomically feasible space. We stress the fact that our method works as a post-processing step in the space of segmentations, making it independent of the predictor, image intensities and modality. We employ denoising autoencoders (DAE) to learn such model.\\

\noindent \textbf{Denoising autoencoders.} DAEs are neural networks designed to reconstruct a clean input from a corrupted version of it \cite{Vincent2010}. In our case, they will be used to reconstruct anatomically plausible segmentation masks from corrupted or erroneous ones. The standard architecture for an autoencoder follows an encoder-decoder scheme (see the Sup. Mat. for a detailed description of the architecture used in this work). The encoder $f_{enc}(S_i)$ is a mapping that transforms the input into a hidden representation $h$. In our case, it consists of successive non-linearities, pooling and convolutional layers, with a final fully connected layer that concentrates all information into a low dimensional code $h$. This code is then feed into the decoder $f_{dec}(h)$, which maps it back to the original input dimensions through a series of up-convolutions and non-linearities. The output of $f_{dec}(h)$ has the same size than the input $S_i$. 

The model is called \textit{denosing} autoenconder because a degradation function $\phi$ is used to degrade the ground-truth segmentation masks, producing noisy segmentations $\hat{S}_i = \phi(S_i)$ used for training. The model is trained to minimize the reconstruction error measured by a loss function based on the Dice coefficient (DSC), a metric used to compare the quality of predicted segmentations with respect to the ground-truth (we refer the reader to \cite{milletari2016v} for a complete description of the Dice loss):
\begin{equation}
    \mathcal{L}_{DAE}(S_i) = DSC(S_i, f_{dec}(f_{enc}(\phi(S_i))).
\end{equation}
The dimensionality of the learned representation $h = f_{enc}(S_i)$ is much lower than the input, producing a bottleneck effect which forces the code $h$ to retain as much information as possible about the input. In that way, minimizing the reconstruction error amounts to maximizing a lower bound on the mutual information between input $S_i$ and the learnt representation $h$ \cite{Vincent2010}.\\

\noindent \textbf{Mask degradation strategy.} The masks used to train the DAE were artificially degraded during training to simulate erroneous segmentations. To this end, we randomly apply the following degradation functions $\phi(S_i)$ to the ground truth masks $S_i$: (i) addition and removal of random geometric shapes (circles, ellipses, lines and rectangles) to simulate over and under segmentations; (ii) morphological operations (e.g. erosion, dilation, etc) with variable kernels to perform more subtle mask modifications and (iii) random swapping of foreground-background labels in the pixels close to the mask borders.\\
 %In addition, data augmentation was performed by randomly changing the size and aspect ratio of the original masks.
 
\noindent \textbf{Post-processing with DAEs.} The proposed method is rooted in the so-called manifold assumption \cite{chapelle2009semi}, which states that natural high dimensional data (like anatomical segmentation masks) concentrate close to a non-linear low-dimensional manifold. We learn such low-dimensional anatomically plausible manifold using the aforementioned DAE. Then, given a segmentation mask $S^P_i$ obtained with an arbitrary predictor $P$ (e.g. CNN or RF), we project it into that manifold using $f_{enc}$ and reconstruct the corresponding anatomically feasible mask with $f_{dec}$. Unlike other methods like \cite{Oktay2017,Ravishankar2017} which incorporate the anatomical priors while training the segmentation network, we choose to make it a post-processing step. In that way, we achieve independence with respect to the initial predictor, and enable improvement for arbitrary segmentation methods.
%Recent studies \cite{pawlowski2018unsupervised,uzunova2019unsupervised} show that different autoencoders trained with healthy brain images (operating in the intensity domain) can be used to perform anomaly detection on pathological brain images, by just looking at the differences between the original pathological image and the one processed by the autoencoder. In the same spirit, 

Our hypothesis (empirically validated by the following experiments) is that those masks which are far from the anatomical space, will be mapped to a similar, but anatomically plausible segmentation. Meanwhile, masks which are anatomically correct, will be mapped to themselves, incurring in almost no modification.

\section{Experiments and Discussion}
\noindent \textbf{Database description.} We benchmark the proposed method in the context of lung segmentation in X-Ray images, using the Japanese Society of Radiological Technology (JSRT) database \cite{JSRT}. JSRT is a public database containing 247 PA chest X-ray images with expert segmentation masks, of 2048x2048 pixels and isotropic spacing of 0.175 mm/pixel, which are downsampled to 1024x1024 in our experiments. Lungs present high variability among subjects, making the representation learning task especially challenging. We divide the database in 3 folds considering 70\% for training, 10\% for validation and 20\% for testing.\\

\noindent \textbf{Post-processing with CRF.}
We compare Post-DAE with the SOA post-processing method based on a fully connected CRF \cite{krahenbuhl2011efficient}. The CRF is used to impose connectivity constraints to a given segmentation, based on the assumption that objects are usually continuous and nearby pixels with similar appearance should be assigned the same object label. We use an efficient implementation of a dense CRF\footnote{We used the public implementation available at \url{https://github.com/lucasb-eyer/pydensecrf} with Potts compatibility function and hand-tuned parameters $\theta_\alpha=17$, $\theta_\beta=3$, $\theta_\gamma=3$ chosen using the validation fold. See the implementation website for more details about the aforementioned parameters.} that can handle large pixel neighbourhoods in reasonable inference times. Differently from our method which uses only binary segmentations for post-processing, the CRF incorporates intensity information from the original images. Therefore, it has to be re-adjusted depending on the image properties of every dataset. Instead, our method is trained once and can be used independently of the image propierties. Note that we do not compare Post-DAE with other methods like \cite{Oktay2017,Ravishankar2017} which incorporate anatomical priors while training the segmentation method itself, since these are not post-processing strategies.\\

\noindent \textbf{Baseline segmentation methods.} We train two different models which produce segmentation masks of various qualities to benchmark our post-processing method. The first model is a CNN based on UNet architecture \cite{RonnebergerUnet15} (see the Sup. Mat. for a detailed description of the architecture and the training parameters such as optimizer, learning rate, etc.). The UNet was implemented in Keras and trained in GPU using a Dice loss function. To evaluate the effect of Post-DAE in different masks, we save the UNet model every 5 epochs during training, and predict segmentation masks for the test fold using all these models. The second method is a RF classifier trained using intensity and texture features. We used Haralick \cite{haralick1973textural} features which are based on gray level co-ocurrency in image patches. We adopted a public implementation available online with default parameters\footnote{The source code and a complete description of the method is publicly available online at: \url{https://github.com/dgriffiths3/ml_segmentation}} which produces acceptable segmentation masks.\\
\begin{figure}[t!]
   \includegraphics[width=\textwidth]{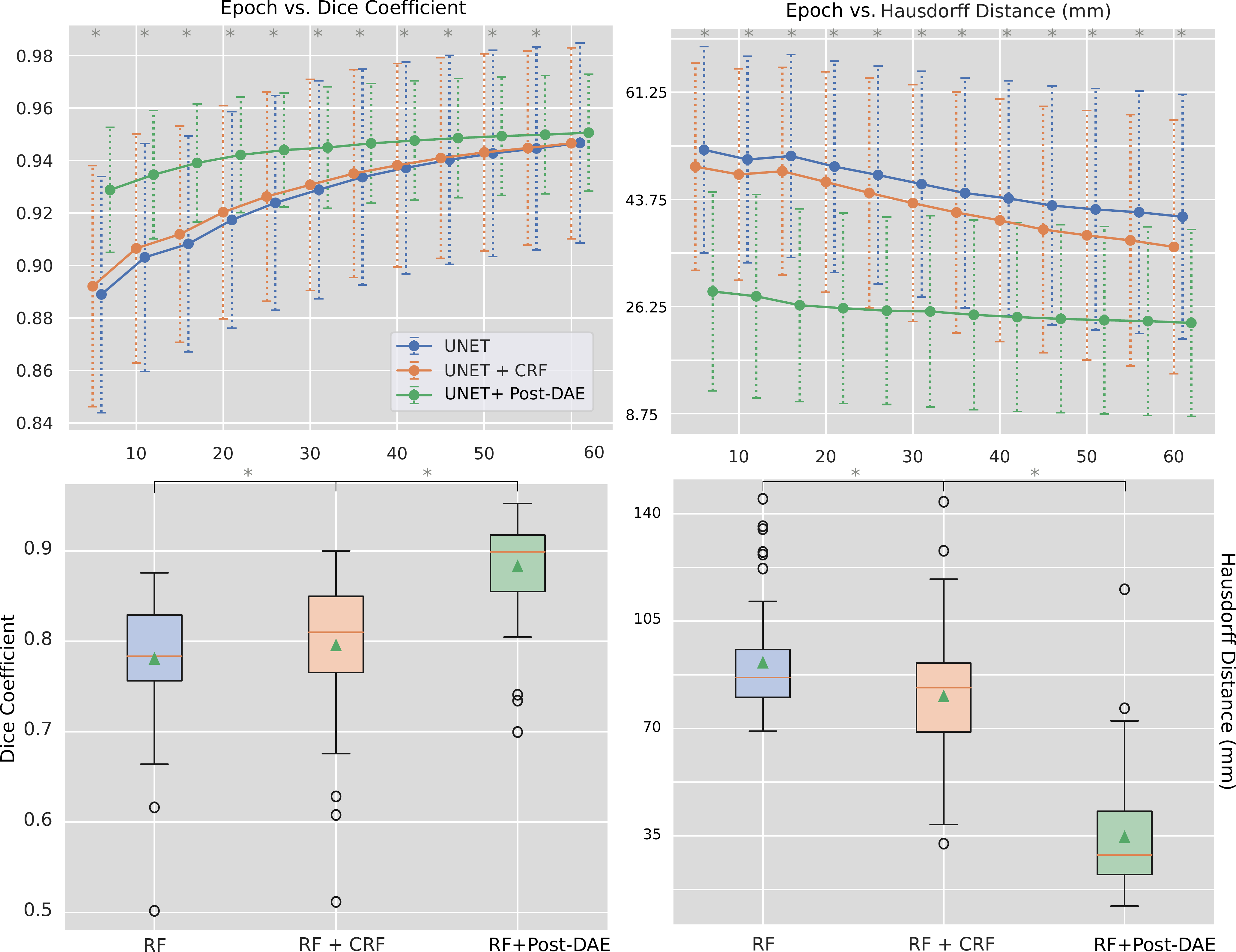}
    \caption{Quantitative evaluation of the proposed method. We compare Post-DAE with the classic fully-connected CRF \cite{krahenbuhl2011efficient} adopted as post-processing step by many segmentation methods like \cite{KamnitsasDeepmedic16}. Top row shows mean and standard deviation for post-processing UNet predictions on the test fold at different training stages (from 5 epochs to convergence). We use Dice coefficient and Hausdorff distance to measure the segmentation quality. Bottom row show results for post-processing the Random Forest predictions. The symbol $*$ indicates that Post-DAE outperforms the other methods (no post-processing and CRF) with statistical significance (p-value $<$ 0.05 according to Wilcoxon test). The green triangle in the box indicates the mean value.}
\label{fig:quantitativeResults}
\end{figure}

\noindent \textbf{Results and discussion.} Figure \ref{fig:qualitativeResults} shows some visual examples while Figure \ref{fig:quantitativeResults} summarizes the quantitative results (see the video in the Sup. Mat. for more visual results). Both figures show the  consistent improvement that can be obtained using Post-DAE as a post-processing step, specially in low quality segmentation masks like those obtained by the RF model and the UNet trained for only 5 epochs. In these cases, substantial improvements are obtained in terms of Dice coefficient and Hausdorff distance, by bringing the erroneous segmentation masks into an anatomically feasible space. In case of segmentations that are already of good quality (like the UNet trained until convergence), the post-processing significantly improves the Hausdorff distance, by erasing spurious segmentations (holes in the lung and small isolated blobs) that remain even in well trained models. When compared with CRF post-processing, Post-DAE significantly outperforms the baseline in the context of anatomical segmentation. In terms of running time, the CRF model takes 1.3 seconds in a Intel i7-7700 CPU, while Post-DAE takes 0.7 seconds in a Titan Xp GPU.

One of the limitations of Post-DAE is related to data regularity. In case of anatomical structures like lung, heart or liver, even if we found high inter-subject variability, the segmentation masks are somehow uniform in terms of shape and topology. Even pathological organs tend to have similar structure, which can be well-encoded by the DAE (specially if pathological cases are seen during training). However, in other cases like brain lesions or tumors where shape is not that regular, it is not clear how Post-DAE would perform. This case lies out of the scope of this paper, but will be explored as future work.\\

\noindent \textbf{Conclusions and future works.}
In this work we have showed, for the first time in the MIC community, that autoencoders can be used as an independent post-processing step to incorporate anatomical priors into arbitrary segmentation methods. Post-DAE can be easily implemented, is fast at inference, can cope with arbitrary shape priors and is independent of the image modality and segmentation method. In the future, we plan to extend this method to muti-class and volumetric segmentation cases (like anatomical segmentation in brain images).

\section{Acknowledgments}
%***************************************************************\\
%***************************************************************\\
%***************************************************************\\
%***************************************************************
EF is beneficiary of an AXA Research Fund grant. The authors gratefully acknowledge NVIDIA Corporation with the donation of the Titan Xp GPU used for this research, and the support of UNL (CAID-PIC-50420150100098LI) and ANPCyT (PICT 2016-0651).

\section{Appendix A: Model details}
\noindent \textbf{UNet details}: The UNet model (see Table \ref{tab:unet}) receives a 1024x1024 gray image as input and was trained using the soft Dice loss \cite{milletari2016v}, batch size of 4, Adam optimizer with learning rate 1e-5 and the other parameters as by Keras default. We also used dropout for regularization, including a dropout layer after layer $L_5$ with keep probability p=0.5.\\

\noindent \textbf{Post-DAE}: Post-DAE (see Table \ref{tab:postdae})receives a 1024x1024 binary segmentation as input. The network was also trained to minimize the Dice loss function using Adam Optimizer. The best performance was achieve with learning rate 0.0001; batch size 15 and 150 epochs.

\begin{table}[t]
\caption{\label{tab:unet} Detailed description of the UNet architecture used as baseline model segmentation·}
\begin{center}
\begin{tabular}{llllll}
\hline
    &                 & \textbf{Kernel}\hspace{0.1 in}                      & \textbf{Stride} \hspace{0.1 in}                      & \#\textbf{Kernels}\hspace{0.1 in}  & \textbf{NonLin}  \\ \hline
   
L1  & Conv            & (f:3,3)                     & (s:1,1)                     & (N:16)    & ReLu    \\
    & Conv            & (f:3,3)                     & (s:1,1)                     & (N:16)    & ReLu    \\ 
    & Max Pooling     & (f:2,2)                     & (s:2,2)                     &           &         \\ \hline
L2  & Conv            & (f:3,3)                     & (s:1,1)                     & (N:32)    & ReLu    \\
    & Conv            & (f:3,3)                     & (s:1,1)                     & (N:32)    & ReLu    \\ 
    & Max Pooling     & (f:2,2)                     & (s:2,2)                     &           &         \\ \hline
L3  & Conv            & (f:3,3)                     & (s:1,1)                     & (N:64)    & ReLu    \\
    & Conv            & (f:3,3)                     & (s:1,1)                     & (N:64)    & ReLu    \\ 
    & Max Pooling     & (f:2,2)                     & (s:2,2)                     &           &         \\ \hline
L4  & Conv            & (f:3,3)                     & (s:1,1)                     & (N:128)   & ReLu    \\
    & Conv            & (f:3,3)                     & (s:1,1)                     & (N:128)   & ReLu    \\ 
    & Max Pooling     & (f:2,2)                     & (s:2,2)                     &           &         \\ \hline
L5  & Conv            & (f:3,3)                     & (s:1,1)                     & (N:256)   & ReLu    \\ 
    & Conv            & (f:3,3)                     & (s:1,1)                     & (N:256)   & ReLu    \\ \hline
L6  & UpConv          & (f:3,3)& (s:1,1) & (N:128)   & ReLu    \\
    & Conv            & (f:3,3) & (s:1,1) & (N:128)   & ReLu    \\ 
    & Conv            & (f:3,3)                     & (s:1,1)                     & (N:128)   & ReLu    \\\hline
L7  & UpConv          & (f:3,3)                     & (s:1,1)                     & (N:64)    & ReLu    \\ 
    & Conv            & (f:3,3)                     & (s:1,1)                     & (N:64)    & ReLu    \\
    & Conv            & (f:3,3)                     & (s:1,1)                     & (N:64)    & ReLu    \\ \hline
L8  & UpConv          & (f:3,3)                     & (s:1,1)                     & (N:32)    & ReLu    \\
    & Conv            & (f:3,3)                     & (s:1,1)                     & (N:32)    & ReLu    \\ 
    & Conv            & (f:3,3)                     & (s:1,1)                     & (N:32)    & ReLu    \\\hline
L9  & UpConv          & (f:3,3)                     & (s:1,1)                     & (N:16)    & ReLu    \\ 
    & Conv            & (f:3,3)                     & (s:1,1)                     & (N:16)    & ReLu    \\
    & Conv            & (f:3,3)                     & (s:1,1)                     & (N:16)    & ReLu    \\ \hline
  L10  & Conv            & (f:3,3)                     & (s:1,1)                     & (N:2)     & ReLu    \\
 & Conv            & (f:1,1)                     & (s:1,1)                     & (N:1)     & Sigmoid \\\hline

\end{tabular}
\end{center}
\end{table}

\begin{table}[t]
\caption{\label{tab:postdae} Detailed architecture of the simple denoising autoencoder model used to implement the proposed Post-DAE.}
\begin{center}
\begin{tabular}{llllll}
\hline
    &        & \textbf{Kernel}                & \textbf{Stride}                & \#\textbf{Kernels} & \textbf{NonLin}  \\ \hline
$L_1$  & Conv   & (f:3,3)               & (s:2,2)               & (N:16)    & ReLu    \\
    & Conv   & (f:3,3)               & (s:1,1)               & (N:16)    & ReLu    \\ \hline
L2  & Conv   & (f:3,3)               & (s:2,2)               & (N:32)    & ReLu    \\
    & Conv   & (f:3,3)               & (s:1,1)               & (N:32)    & ReLu    \\ \hline
L3  & Conv   & (f:3,3)               & (s:2,2)               & (N:32)    & ReLu    \\
    & Conv   & (f:3,3)               & (s:1,1)               & (N:32)    & ReLu    \\ \hline
L4  & Conv   & (f:3,3)               & (s:2,2)               & (N:32)    & ReLu    \\
    & Conv   & (f:3,3)               & (s:1,1)               & (N:32)    & ReLu    \\ \hline
L5  & Conv   & (f:3,3)               & (s:2,2)               & (N:32)    & ReLu    \\ \hline
L6  & FC     & \multicolumn{1}{c}{-} & \multicolumn{1}{c}{-} & (N:512)   & None    \\ \hline
L6  & FC     & \multicolumn{1}{c}{-} & \multicolumn{1}{c}{-} & (N:1024)  & Relu    \\ \hline
L8  & UpConv & (f:3,3)               & (s:1,1)               & (N:16)    & ReLu    \\
    & Conv   & (f:3,3)               & (s:1,1)               & (N:16)    & ReLu    \\ \hline
L9  & UpConv & (f:3,3)               & (s:1,1)               & (N:16)    & ReLu    \\
    & Conv   & (f:3,3)               & (s:1,1)               & (N:16)    & ReLu    \\ \hline
L10 & UpConv & (f:3,3)               & (s:1,1)               & (N:16)    & ReLu    \\
    & Conv   & (f:3,3)               & (s:1,1)               & (N:16)    & ReLu    \\ \hline
L11 & UpConv & (f:3,3)               & (s:1,1)               & (N:16)    & ReLu    \\
    & Conv   & (f:3,3)               & (s:1,1)               & (N:16)    & ReLu    \\ \hline
L12 & UpConv & (f:3,3)               & (s:1,1)               & (N:16)    & ReLu    \\
    & Conv   & (f:3,3)               & (s:1,1)               & (N:1)     & Sigmoid \\ \hline
\end{tabular}
\end{center}
\end{table}

\newpage
\bibliographystyle{splncs}
\bibliography{library}

\end{document}